\documentclass[sn-mathphys-num]{sn-jnl}% Math and Physical Sciences Numbered Reference Style
\usepackage{graphicx}%
\usepackage{multirow}%
\usepackage{amsmath,amssymb,amsfonts}%
\usepackage{amsthm}%
\usepackage{mathrsfs}%
\usepackage[title]{appendix}%
\usepackage{xcolor}%
\usepackage{textcomp}%
\usepackage{manyfoot}%
\usepackage{booktabs}%
\usepackage{algorithm}%
\usepackage{algorithmicx}%
\usepackage{algpseudocode}%
\usepackage{listings}%
\usepackage{rotating}  
\usepackage{dcolumn}       % <--- added
\usepackage{tablefootnote}
\theoremstyle{thmstyleone}%
\newtheorem{theorem}{Theorem}%  meant for continuous numbers
\newtheorem{proposition}[theorem]{Proposition}% 

\theoremstyle{thmstyleone}%
\newtheorem{corollary}{Corollary}

\theoremstyle{thmstyletwo}%
\newtheorem{example}{Example}%
\newtheorem{remark}{Remark}%

\theoremstyle{thmstylethree}%
\newtheorem{definition}{Definition}%

\begin{document}

\title[Article Title]{Absorption and Inertness in Coarse-Grained Arithmetic: A Heuristic Application to the St. Petersburg Paradox}


\author*[1]{\fnm{Takashi} \sur{Izumo}}\email{izumo.takashi@nihon-u.ac.jp}

\affil*[1]{\orgdiv{College of Law}, \orgname{Nihon University}, \orgaddress{\street{Kandamisakicho}, \city{Chiyodaku}, \postcode{1018375}, \state{Tokyo}, \country{Japan}, ORCID: 0000-0003-0008-4729}}


\abstract{
The St. Petersburg paradox presents a longstanding challenge in decision theory:
its classical expected value diverges, yet no correspondingly large finite stake is
typically regarded as rational. Traditional responses introduce auxiliary assumptions,
such as diminishing marginal utility, temporal discounting, or extended number
systems. This paper explores a different approach based on a modified operation
of addition defined over coarse-grained partitions of the underlying numerical scale.
In this framework, exact values are grouped into ordered grains, each grain is assigned
an internal representative, and addition proceeds by repeated projection to those
representatives. On this basis, the paper defines coarse representative addition and
coarse cell addition, and studies several of their structural properties, including
absorption, inertness, and non-associativity. In particular, repeated additions may
eventually cease to change the coarse state, a phenomenon called inertness.
The paper then applies this framework heuristically to the St. Petersburg setting
by considering a rescaled sequence corresponding to its equal expected increments,
and shows that this sequence can become inert under a suitably chosen countable
partition and representative map. The claim is not that the paradox is resolved
within standard decision theory, nor that the classical expectation becomes finite in
the ordinary probabilistic sense. Rather, the contribution is structural and heuristic:
it exhibits an explicit mathematical mechanism through which a divergent reward
structure may fail to produce unbounded growth once aggregation itself is made coarse.
More broadly, the framework may be relevant to the study of bounded numerical
cognition and behavioral models of aggregation.
}

\keywords{St. Petersburg Paradox, Infinite Game, Coarse-Graining, Coarse-Grained Partitions, Coarse Addition}



\maketitle

\section{Introduction}\label{sec1}

\subsection{Background}\label{sec1.1}

In games whose expected payoff diverges to infinity, players cannot set a rational finite entry fee. The St.\ Petersburg paradox \cite{Bernoulli1738} exemplifies this: you toss a fair coin until it comes up tails on the \(n\)th toss, and you receive \(2^{n-1}\) units. Since
\[
\sum_{n=1}^\infty \frac{1}{2^n}\,2^{n-1}
=
\sum_{n=1}^\infty \frac{1}{2}
=
+\infty,
\]
there is no finite fee equal to that infinite expectation.

Some of the main proposed solutions in the existing literature are as follows.

\begin{enumerate}
\item Classical economics addresses the St.\ Petersburg paradox by replacing monetary payoffs with a concave utility function \cite{Bernoulli1738}. Under this approach, the growth of utility is slower than the growth of payoffs, so that an infinite monetary expectation need not translate into an infinite expected utility. The force of the solution, however, depends on the particular curvature of the utility function and on how rapidly the payoffs increase.

\item Temporal discounting handles infinite payoff streams by multiplying each term by \(\delta^n\) with \(0<\delta<1\) \cite{Samuelson1937}. This gives progressively smaller weight to later payoffs and thereby turns many divergent series into convergent ones. At the same time, the choice of \(\delta\) is not uniquely determined, and the interpretation of discounting often combines attitudes toward time with attitudes toward risk and evaluation.

\item Recent unbounded-utility decision theory retains infinite expectations and instead studies how they may be compared within richer ordering frameworks \cite{Goodsell2024}. On this view, the problem is not necessarily to eliminate divergence, but to determine how divergent or unbounded expectations can still be handled in a coherent decision-theoretic manner. Such approaches therefore reconsider the structure of preference or comparison rather than forcing ordinary convergence.

\item Prospect theory explains many behavioral-economic anomalies by combining a value function with probability weighting \cite{KahnemanTversky1979}. This allows the evaluation of risky prospects to depart systematically from classical expected-utility theory. However, even after such transformations, a series of the form \(\sum v(x_n)\) may still diverge if the underlying payoffs grow sufficiently quickly.
\end{enumerate}

This paper takes a different tack. Rather than modifying utility, introducing a discount factor, or extending the number system, it studies a new aggregation rule called \emph{coarse addition}. Under this rule, values are grouped into finite grains, each grain is assigned a representative, and addition proceeds through repeated projection to those representatives. As a result, a running total that would diverge under ordinary addition may, under a suitable coarse-grained structure, display eventual inertness after finitely many steps.

The chief aim of this paper is not to provide a final, once-and-for-all solution to the St.\ Petersburg paradox. Rather, it asks how a \emph{coarse-grained} cognitive system might process a divergent reward structure in the first place. If human judgment is often coarse rather than fully fine-grained, then models of evaluation and aggregation may also need to incorporate a compatible level of coarseness. More generally, coarse-graining---the replacement of fine-grained detail by higher-level aggregates---appears across a wide range of disciplines \cite{Marrink2004,Marrink2007,souza2021,Baschnagel2000,Reith2003}. Related coarse-grained perspectives have also been explored in game theory and economics \cite{izumo2025coarsegrainedgamesframeworkbounded}.

The remainder of this paper is organized as follows. Section~\ref{sec2} introduces the basic setup of coarse-grained arithmetic, including countable interval partitions, cell maps, internal representative maps, and the induced operations of coarse representative addition and coarse cell addition. Section~\ref{sec3} then studies absorption and provides its numerical characterization together with a sufficient condition in terms of the absorption margin. Section~\ref{sec4} turns to inertness, defines left-associative coarse partial sums, and establishes sufficient conditions under which repeated additions become eventually constant. Section~\ref{sec5} examines the non-associativity of coarse addition, while also identifying a special case in which associativity is recovered. Section~\ref{sec6} applies the framework heuristically to the St.\ Petersburg paradox by showing that a rescaled constant increment sequence can become inert under a suitable partition and representative map. Section~\ref{sec7} concludes with a summary of the main results, limitations, and directions for future research.

\section{Setup}\label{sec2}

\subsection{Underlying Scale and Countable Interval Partitions}

Throughout this paper, we work on the discrete totally ordered set
\[
U := \mathbb{N}_0 = \{0,1,2,\dots\}
\]
equipped with the natural order~$\leq$.

Our aim is to study a coarse-grained form of addition on $U$. To this end, we first
introduce a countable partition of $U$ into finite ordered blocks.

\begin{definition}[Countable discrete coarse-grained partition]
A \emph{countable discrete coarse-grained partition} of $U$ is a family
\[
\pi = \{G_{\pi,i}\}_{i\in I_\pi}
\]
satisfying the following conditions:
\begin{enumerate}
    \item $I_\pi$ is a countable totally ordered index set;
    \item for each $i\in I_\pi$, the set $G_{\pi,i}\subseteq U$ is a nonempty finite interval;
    \item the family is pairwise disjoint, i.e.,
    \[
    G_{\pi,i}\cap G_{\pi,j}=\varnothing \qquad (i\neq j);
    \]
    \item the family covers $U$, i.e.,
    \[
    U=\bigsqcup_{i\in I_\pi} G_{\pi,i};
    \]
    \item the indexing respects the order on $U$, i.e., whenever $i<j$ in $I_\pi$,
    \[
    (\forall x\in G_{\pi,i})(\forall y\in G_{\pi,j})\; x<y.
    \]
\end{enumerate}
Each set $G_{\pi,i}$ is called a \emph{grain} of the partition~$\pi$.
\end{definition}

Since every grain is a finite interval in $\mathbb{N}_0$, we may write
\[
G_{\pi,i}=\{a_i,a_i+1,\dots,b_i\}
\]
for some integers $a_i\leq b_i$, where $|G_{\pi,i}|=b_i-a_i+1$.

\begin{example}
A typical example is given by
\[
G_{\pi,1}=\{0\},\qquad
G_{\pi,2}=\{1\},\qquad
G_{\pi,3}=\{2,3\},\qquad
G_{\pi,4}=\{4,5,6\},
\]
\[
G_{\pi,5}=\{7,8,9,10,11\},\qquad
G_{\pi,6}=\{12,13,14,15,16,17,18,19\},\ \dots
\]
More generally, one may construct $\pi$ by specifying the size of each grain and placing
the grains consecutively along~$U$.
\end{example}

\subsection{Cell Map and Internal Representative Map}

Given a countable discrete coarse-grained partition $\pi=\{G_{\pi,i}\}_{i\in I_\pi}$,
each element of $U$ belongs to exactly one grain. Hence we obtain the canonical
cell map.

\begin{definition}[Cell map]
Let $\pi=\{G_{\pi,i}\}_{i\in I_\pi}$ be a countable discrete coarse-grained partition of~$U$.
The \emph{cell map}
\[
\psi_\pi:U\to \pi
\]
is defined by
\[
\psi_\pi(x)=G_{\pi,i}\quad \Longleftrightarrow \quad x\in G_{\pi,i}.
\]
\end{definition}

To define coarse addition, we also need to assign to each grain a single representative
value. In order to preserve the internal coherence of the partition structure, we require
that the representative be chosen from the grain itself.

\begin{definition}[Internal representative map]
Let $\pi=\{G_{\pi,i}\}_{i\in I_\pi}$ be a countable discrete coarse-grained partition of~$U$.
An \emph{internal representative map} for $\pi$ is a map
\[
\phi:\pi\to U
\]
such that
\[
\phi(G_{\pi,i})\in G_{\pi,i}
\qquad\text{for every }i\in I_\pi.
\]
\end{definition}

The requirement $\phi(G_{\pi,i})\in G_{\pi,i}$ ensures that the representative value is
genuinely selected from the original grain, rather than imported from an external domain.

\begin{example}[Typical choices of representatives]
Let
\[
G_{\pi,i}=\{u_1<u_2<\dots<u_n\}.
\]
Typical internal representative maps include:
\begin{itemize}
    \item the minimum representative,
    \[
    \phi_{\min}(G_{\pi,i}) := u_1;
    \]
    \item the maximum representative,
    \[
    \phi_{\max}(G_{\pi,i}) := u_n;
    \]
    \item the lower median representative,
    \[
    \phi_{\mathrm{lmed}}(G_{\pi,i}) := u_{\lfloor (n+1)/2 \rfloor};
    \]
    \item the upper median representative,
    \[
    \phi_{\mathrm{umed}}(G_{\pi,i}) := u_{\lceil (n+1)/2 \rceil}.
    \]
\end{itemize}
In particular, when the underlying scale is~$\mathbb{N}_0$, both the lower and upper median representatives are well defined and always belong to the grain, even when $|G_{\pi,i}|$ is even.
\end{example}

\subsection{Coarse Addition}

We now define two coarse-grained addition operations: one on elements of~$U$ and the
other on grains of the partition.

\begin{definition}[Coarse representative addition]
Let $\pi=\{G_{\pi,i}\}_{i\in I_\pi}$ be a countable discrete coarse-grained partition of~$U$,
and let $\phi$ be an internal representative map for~$\pi$.
The \emph{coarse representative addition}
\[
\oplus_{\pi,\phi}:U\times U\to U
\]
is defined by
\[
x\oplus_{\pi,\phi} y
:=
\phi\!\left(
\psi_\pi\bigl(\phi(\psi_\pi(x))+\phi(\psi_\pi(y))\bigr)
\right).
\]
\end{definition}

This operation proceeds in four steps:
\begin{enumerate}
    \item map $x$ and $y$ to their grains;
    \item replace each grain by its representative;
    \item add the two representatives in the ordinary sense;
    \item map the resulting sum back to its grain and select the representative again.
\end{enumerate}

\begin{definition}[Coarse cell addition]
Let $\pi=\{G_{\pi,i}\}_{i\in I_\pi}$ be a countable discrete coarse-grained partition of~$U$,
and let $\phi$ be an internal representative map for~$\pi$.
The \emph{coarse cell addition}
\[
\boxplus_{\pi,\phi}:\pi\times\pi\to \pi
\]
is defined by
\[
G_{\pi,i}\boxplus_{\pi,\phi} G_{\pi,j}
:=
\psi_\pi\bigl(\phi(G_{\pi,i})+\phi(G_{\pi,j})\bigr).
\]
\end{definition}

The two operations are compatible in the obvious sense:
if $x\in G_{\pi,i}$ and $y\in G_{\pi,j}$, then
\[
x\oplus_{\pi,\phi} y
=
\phi\bigl(G_{\pi,i}\boxplus_{\pi,\phi} G_{\pi,j}\bigr).
\]

\section{Absorption}\label{sec3}

\subsection{Basic Definition}

We now formalize the central phenomenon of this paper: under coarse addition,
a grain may absorb another grain, in the sense that adding the second grain
does not move the result outside the first one.

\begin{definition}[Absorption of grains]
Let $\pi=\{G_{\pi,i}\}_{i\in I_\pi}$ be a countable discrete coarse-grained partition of
$U=\mathbb N_0$, and let $\phi$ be an internal representative map for~$\pi$.

For two grains $G,H\in \pi$, we say that \emph{$G$ absorbs $H$} if
\[
G \boxplus_{\pi,\phi} H = G.
\]
Equivalently, we also say that \emph{$H$ is absorbed by $G$}.
\end{definition}

Thus, absorption means that the representative of $H$, when added to the
representative of $G$, still lands in the same grain $G$ after coarse remapping.

\begin{remark}
In general, absorption is not symmetric. That is, it may happen that
\[
G \boxplus_{\pi,\phi} H = G
\qquad\text{but}\qquad
H \boxplus_{\pi,\phi} G \neq H.
\]
Hence absorption should not be confused with an equivalence relation.
\end{remark}

\subsection{Numerical Characterization}

Since the grains are finite intervals in $\mathbb N_0$, absorption can be characterized
numerically.

\begin{proposition}[Characterization of absorption]
Let
\[
G=\{a,a+1,\dots,b\}\in \pi,\qquad H\in \pi.
\]
Then $G$ absorbs $H$ if and only if
\[
\phi(G)+\phi(H)\in G.
\]
Equivalently,
\[
G \boxplus_{\pi,\phi} H = G
\quad\Longleftrightarrow\quad
a \le \phi(G)+\phi(H)\le b.
\]
\end{proposition}

\begin{proof}
By definition,
\[
G \boxplus_{\pi,\phi} H
=
\psi_\pi\bigl(\phi(G)+\phi(H)\bigr).
\]
Hence
\[
G \boxplus_{\pi,\phi} H = G
\]
holds if and only if the ordinary sum $\phi(G)+\phi(H)$ belongs to the grain $G$.
Since $G=\{a,a+1,\dots,b\}$ is an interval in $\mathbb N_0$, this is equivalent to
\[
a \le \phi(G)+\phi(H)\le b.
\]
\end{proof}

\subsection{Absorption Margin and a Sufficient Condition}

The upper room left above the representative of a grain determines whether small
increments can be absorbed.

Recall that for a grain
\[
G=\{a,a+1,\dots,b\},
\]
its absorption margin is
\[
\mu_\phi(G):=b-\phi(G).
\]

The following proposition shows that absorption is controlled by comparing
the representative of the increment grain with the absorption margin of the
candidate absorbing grain.

\begin{proposition}[Absorption criterion via margin]
Let
\[
G=\{a,a+1,\dots,b\}\in \pi,\qquad H\in \pi.
\]
If
\[
\phi(H)\le \mu_\phi(G)=b-\phi(G),
\]
then
\[
\phi(G)+\phi(H)\le b.
\]
Hence, if in addition
\[
\phi(G)+\phi(H)\ge a,
\]
we obtain
\[
a\le \phi(G)+\phi(H)\le b,
\]
and therefore
\[
G\boxplus_{\pi,\phi}H=G.
\]
For $U=\mathbb N_0$, the lower bound is automatic because
\[
\phi(G)\ge a \quad\text{and}\quad \phi(H)\ge 0.
\]
Thus, in this setting,
\[
\phi(H)\le \mu_\phi(G)\implies G\boxplus_{\pi,\phi}H=G.
\]
\end{proposition}

\begin{proof}
By assumption,
\[
\phi(H)\le \mu_\phi(G)=b-\phi(G).
\]
Hence
\[
\phi(G)+\phi(H)\le b.
\]
Also, since $\phi(G)\in G$, we have $\phi(G)\ge a$, and since $\phi(H)\in\mathbb N_0$,
\[
\phi(G)+\phi(H)\ge \phi(G)\ge a.
\]
Therefore
\[
a\le \phi(G)+\phi(H)\le b.
\]
By the previous proposition, this implies
\[
G \boxplus_{\pi,\phi} H = G.
\]
\end{proof}

\begin{corollary}[Self-absorption criterion]
Let $G\in \pi$. If
\[
\phi(G)\le \mu_\phi(G),
\]
then $G$ absorbs itself:
\[
G \boxplus_{\pi,\phi} G = G.
\]
\end{corollary}

\begin{proof}
Apply the previous proposition with $H=G$.
\end{proof}

\subsection{Example}

\begin{example}
Let
\[
G_{\pi,1}=\{0,1,2\},\qquad
G_{\pi,2}=\{3,4,5\},\qquad
G_{\pi,3}=\{6,7,8,9,10,11,12,13,14,15,16\},
\]
and let $\phi=\phi_{\mathrm{lmed}}$ be the lower median representative. Then
\[
\phi(G_{\pi,2})=4,\qquad \phi(G_{\pi,3})=11.
\]
Since
\[
\mu_\phi(G_{\pi,3})=16-11=5
\]
and
\[
\phi(G_{\pi,2})=4\le 5,
\]
it follows that $G_{\pi,3}$ absorbs $G_{\pi,2}$:
\[
G_{\pi,3}\boxplus_{\pi,\phi} G_{\pi,2}=G_{\pi,3}.
\]
Indeed,
\[
\phi(G_{\pi,3})+\phi(G_{\pi,2})=11+4=15\in G_{\pi,3}.
\]
\end{example}

\section{Inertness}\label{sec4}

\subsection{Left-Associative Coarse Partial Sums}

We now turn from single instances of absorption to the behavior of infinite
coarse addition processes. Since coarse addition is generally non-associative,
we must fix a bracketing convention. Throughout this paper, all infinite coarse
sums are understood in the left-associative sense.

\begin{definition}[Left-associative coarse partial sums on grains]
Let $\pi=\{G_{\pi,i}\}_{i\in I_\pi}$ be a countable discrete coarse-grained partition
of $U=\mathbb N_0$, and let $\phi$ be an internal representative map for~$\pi$.

Given a sequence of grains
\[
(H_n)_{n\in\mathbb N}\subseteq \pi,
\]
define the sequence of \emph{coarse partial sums on grains}
\[
(S_n^{\mathrm{cell}})_{n\in\mathbb N}\subseteq \pi
\]
recursively by
\[
S_1^{\mathrm{cell}} := H_1,
\qquad
S_n^{\mathrm{cell}} := S_{n-1}^{\mathrm{cell}} \boxplus_{\pi,\phi} H_n
\quad (n\ge 2).
\]
\end{definition}

\begin{definition}[Left-associative coarse partial sums on elements]
Let
\[
(x_n)_{n\in\mathbb N}\subseteq U.
\]
Define the sequence of \emph{coarse partial sums on elements}
\[
(S_n^{\mathrm{rep}})_{n\in\mathbb N}\subseteq U
\]
recursively by
\[
S_1^{\mathrm{rep}} := x_1,
\qquad
S_n^{\mathrm{rep}} := S_{n-1}^{\mathrm{rep}} \oplus_{\pi,\phi} x_n
\quad (n\ge 2).
\]
\end{definition}

The two notions are compatible in the obvious way: if
\[
H_n=\psi_\pi(x_n)\qquad (n\in\mathbb N),
\]
then
\[
S_n^{\mathrm{rep}}=\phi\!\left(S_n^{\mathrm{cell}}\right)
\qquad (n\in\mathbb N).
\]

\subsection{Inertness on the Grain Level}

The central phenomenon is that the coarse partial sums may eventually stabilize
at a single grain, even though the corresponding ordinary sum would continue
to grow.

\begin{definition}[Inertness on the grain level]
Let $(H_n)_{n\in\mathbb N}\subseteq \pi$ be a sequence of grains, and let
$(S_n^{\mathrm{cell}})_{n\in\mathbb N}$ be its left-associative coarse partial sums.

We say that the sequence $(H_n)$ is \emph{inert at the grain} $G^*\in\pi$
if there exists $N\in\mathbb N$ such that
\[
S_n^{\mathrm{cell}} = G^*
\qquad\text{for all }n\ge N.
\]
In this case, we also say that the coarse addition process exhibits
\emph{grain-level inertness} at~$G^*$.
\end{definition}

Thus, grain-level inertness means eventual stabilization of the coarse partial
sums in the partition space~$\pi$.

\subsection{Inertness on the Representative Level}

Since each grain carries a representative value, the stabilization of grains
immediately induces stabilization of representative values.

\begin{definition}[Inertness on the representative level]
Let $(x_n)_{n\in\mathbb N}\subseteq U$ be a sequence of elements, and let
$(S_n^{\mathrm{rep}})_{n\in\mathbb N}$ be its left-associative coarse partial sums.

We say that the sequence $(x_n)$ is \emph{inert at the value} $u^*\in U$
if there exists $N\in\mathbb N$ such that
\[
S_n^{\mathrm{rep}} = u^*
\qquad\text{for all }n\ge N.
\]
\end{definition}

\begin{remark}
If $(H_n)$ is inert at a grain $G^*$, then the induced representative partial sums
are inert at the value $\phi(G^*)$. In this sense, representative-level inertness
is a direct consequence of grain-level inertness.
\end{remark}

\subsection{A Sufficient Condition for Inertness}

We now show that repeated absorption yields inertness. This gives a simple and
useful sufficient condition.

\begin{proposition}[Eventual repeated absorption implies inertness]
Let $(H_n)_{n\in\mathbb N}\subseteq \pi$ be a sequence of grains. Suppose that
there exist $N\in\mathbb N$ and a grain $G^*\in\pi$ such that
\[
S_N^{\mathrm{cell}} = G^*
\]
and
\[
G^* \boxplus_{\pi,\phi} H_n = G^*
\qquad\text{for all }n\ge N+1.
\]
Then $(H_n)$ is inert at~$G^*$.
\end{proposition}

\begin{proof}
We prove by induction that
\[
S_n^{\mathrm{cell}} = G^*
\qquad\text{for all }n\ge N.
\]

The case $n=N$ holds by assumption. Now assume that
\[
S_n^{\mathrm{cell}} = G^*
\]
for some $n\ge N$. Then
\[
S_{n+1}^{\mathrm{cell}}
=
S_n^{\mathrm{cell}} \boxplus_{\pi,\phi} H_{n+1}
=
G^* \boxplus_{\pi,\phi} H_{n+1}
=
G^*
\]
by the absorption assumption. Hence the claim follows for all $n\ge N$.
\end{proof}

Using the absorption criterion from the previous section, we immediately obtain
a numerical sufficient condition.

\begin{theorem}[Margin-based sufficient condition for inertness]
Let $(H_n)_{n\in\mathbb N}\subseteq \pi$ be a sequence of grains. Suppose that
there exist $N\in\mathbb N$ and a grain $G^*\in\pi$ such that
\[
S_N^{\mathrm{cell}} = G^*
\]
and
\[
\phi(H_n)\le \mu_\phi(G^*)
\qquad\text{for all }n\ge N+1.
\]
Then $(H_n)$ is inert at~$G^*$.
\end{theorem}

\begin{proof}
For every $n\ge N+1$, the condition
\[
\phi(H_n)\le \mu_\phi(G^*)
\]
implies, by the absorption criterion via margin, that
\[
G^* \boxplus_{\pi,\phi} H_n = G^*.
\]
Hence the previous proposition applies.
\end{proof}

\subsection{Representative Version}

The previous theorem yields an immediate representative-level counterpart.

\begin{corollary}[Representative-level inertness]
Let $(x_n)_{n\in\mathbb N}\subseteq U$ be a sequence of elements, and put
\[
H_n:=\psi_\pi(x_n)\qquad (n\in\mathbb N).
\]
Suppose that there exist $N\in\mathbb N$ and a grain $G^*\in\pi$ such that
\[
\psi_\pi(S_N^{\mathrm{rep}})=G^*
\]
and
\[
\phi(H_n)\le \mu_\phi(G^*)
\qquad\text{for all }n\ge N+1.
\]
Then
\[
S_n^{\mathrm{rep}}=\phi(G^*)
\qquad\text{for all }n\ge N.
\]
In particular, $(x_n)$ is inert at the value $\phi(G^*)$.
\end{corollary}

\begin{proof}
Since
\[
\psi_\pi(S_N^{\mathrm{rep}})=G^*,
\]
the corresponding grain-level partial sum at step $N$ is $G^*$. By the previous theorem,
the grain-level process is inert at $G^*$. Therefore, for all $n\ge N$,
\[
S_n^{\mathrm{rep}}=\phi(S_n^{\mathrm{cell}})=\phi(G^*).
\]
\end{proof}

\subsection{Example}

\begin{example}
Let
\[
G_{\pi,1}=\{0,1,2\},\qquad
G_{\pi,2}=\{3,4,5\},\qquad
G_{\pi,3}=\{6,7,8,9,10,11,12,13,14,15,16\},
\]
and let $\phi=\phi_{\mathrm{lmed}}$ be the lower median representative. Then
\[
\phi(G_{\pi,2})=4,\qquad
\phi(G_{\pi,3})=11,\qquad
\mu_\phi(G_{\pi,3})=16-11=5.
\]

Consider the constant sequence
\[
H_n=G_{\pi,2}
\qquad (n\in\mathbb N).
\]
Its partial sums satisfy
\[
S_1^{\mathrm{cell}}=G_{\pi,2},
\qquad
S_2^{\mathrm{cell}}=G_{\pi,2}\boxplus_{\pi,\phi} G_{\pi,2}
=\psi_\pi(4+4)=\psi_\pi(8)=G_{\pi,3}.
\]
Since
\[
\phi(G_{\pi,2})=4\le 5=\mu_\phi(G_{\pi,3}),
\]
the grain $G_{\pi,3}$ absorbs $G_{\pi,2}$. Therefore
\[
S_n^{\mathrm{cell}}=G_{\pi,3}
\qquad\text{for all }n\ge 2.
\]
Hence the sequence is inert at the grain $G_{\pi,3}$, and the corresponding
representative-level process is inert at the value
\[
\phi(G_{\pi,3})=11.
\]
\end{example}

\begin{remark}[Inertness is not automatic]
Not every coarse addition process becomes inert. For some choices of partition and representative map, successive increments continue to push the coarse partial sums into new grains, and no eventual stabilization occurs. Hence inertness is a distinctive possible behavior of coarse addition, not an unavoidable one.
\end{remark}

\section{Non-Associativity}\label{sec5}

A distinctive feature of coarse addition is that it is generally non-associative.
This is a direct consequence of the fact that, after each binary operation, the
intermediate result is projected back to a representative value of a grain. Once
such a projection occurs, information about the fine-grained magnitude is lost,
and the outcome of further additions may depend on the bracketing order.

\subsection{Non-Associativity in General}

\begin{proposition}[Non-associativity may occur]
There exist a countable discrete coarse-grained partition $\pi$ of $U=\mathbb N_0$,
an internal representative map $\phi$, and elements $x,y,z\in U$ such that
\[
(x\oplus_{\pi,\phi} y)\oplus_{\pi,\phi} z
\neq
x\oplus_{\pi,\phi} (y\oplus_{\pi,\phi} z).
\]
Likewise, there exist grains $G,H,K\in\pi$ such that
\[
(G\boxplus_{\pi,\phi} H)\boxplus_{\pi,\phi} K
\neq
G\boxplus_{\pi,\phi} (H\boxplus_{\pi,\phi} K).
\]
\end{proposition}

\begin{proof}
It suffices to provide a counterexample; see Example~\ref{ex:nonassoc}.
\end{proof}

\begin{example}[A counterexample to associativity]
\label{ex:nonassoc}
Consider the partition
\[
G_{\pi,1}=\{0\},\qquad
G_{\pi,2}=\{1\},\qquad
G_{\pi,3}=\{2,3\},\qquad
G_{\pi,4}=\{4,5,6\},\qquad
G_{\pi,5}=\{7,8,9,10,11\},
\]
\[
G_{\pi,6}=\{12,13,14,15,16,17,18,19\}.
\]
Let $\phi=\phi_{\mathrm{lmed}}$ be the lower median representative. Then
\[
\phi(G_{\pi,3})=2,\qquad
\phi(G_{\pi,5})=9,\qquad
\phi(G_{\pi,6})=15.
\]

Take
\[
x=3,\qquad y=3,\qquad z=10.
\]
Since $3\in G_{\pi,3}$ and $10\in G_{\pi,5}$, we compute:
\[
x\oplus_{\pi,\phi} y
=
3\oplus_{\pi,\phi} 3
=
\phi\!\left(\psi_\pi(2+2)\right)
=
\phi(G_{\pi,4})
=
5.
\]
Hence
\[
(3\oplus_{\pi,\phi} 3)\oplus_{\pi,\phi} 10
=
5\oplus_{\pi,\phi} 10
=
\phi\!\left(\psi_\pi(5+9)\right)
=
\phi(G_{\pi,6})
=
15.
\]

On the other hand,
\[
3\oplus_{\pi,\phi} 10
=
\phi\!\left(\psi_\pi(2+9)\right)
=
\phi(G_{\pi,5})
=
9,
\]
so
\[
3\oplus_{\pi,\phi} (3\oplus_{\pi,\phi} 10)
=
3\oplus_{\pi,\phi} 9
=
\phi\!\left(\psi_\pi(2+9)\right)
=
\phi(G_{\pi,5})
=
9.
\]
Therefore
\[
(3\oplus_{\pi,\phi} 3)\oplus_{\pi,\phi} 10 = 15
\neq
9 = 3\oplus_{\pi,\phi} (3\oplus_{\pi,\phi} 10).
\]

The corresponding grain-level computation is:
\[
G_{\pi,3}\boxplus_{\pi,\phi} G_{\pi,3}
=
\psi_\pi(2+2)
=
G_{\pi,4},
\]
hence
\[
(G_{\pi,3}\boxplus_{\pi,\phi} G_{\pi,3})\boxplus_{\pi,\phi} G_{\pi,5}
=
G_{\pi,4}\boxplus_{\pi,\phi} G_{\pi,5}
=
\psi_\pi(5+9)
=
G_{\pi,6}.
\]
By contrast,
\[
G_{\pi,3}\boxplus_{\pi,\phi} G_{\pi,5}
=
\psi_\pi(2+9)
=
G_{\pi,5},
\]
so
\[
G_{\pi,3}\boxplus_{\pi,\phi}(G_{\pi,3}\boxplus_{\pi,\phi} G_{\pi,5})
=
G_{\pi,3}\boxplus_{\pi,\phi} G_{\pi,5}
=
G_{\pi,5}.
\]
Thus
\[
(G_{\pi,3}\boxplus_{\pi,\phi} G_{\pi,3})\boxplus_{\pi,\phi} G_{\pi,5}
=
G_{\pi,6}
\neq
G_{\pi,5}
=
G_{\pi,3}\boxplus_{\pi,\phi}(G_{\pi,3}\boxplus_{\pi,\phi} G_{\pi,5}).
\]
\end{example}

The reason for this failure is structural. In ordinary addition on $\mathbb N_0$,
intermediate sums are preserved exactly. By contrast, coarse addition replaces each
intermediate sum by the representative of the grain into which it falls. This
replacement is generally lossy. As a result, different bracketings may produce
different intermediate representatives, and hence different final outputs.

In particular, non-associativity is closely related to the presence of absorption:
once an intermediate result falls into a grain that absorbs a later increment,
further growth may stop along one bracketing, while a different bracketing may
avoid that absorption and continue into a higher grain.

\subsection{A Special Associative Case}

Although coarse addition is generally non-associative, associativity may hold under
special structural conditions.

\begin{proposition}[A special associative case]
Suppose that $\pi=\{G_{\pi,i}\}_{i\ge 1}$ is a partition of $U=\mathbb N_0$ into
consecutive intervals of constant odd width
\[
w=2m+1
\qquad (m\in\mathbb N_0),
\]
namely
\[
G_{\pi,i}
=
\{w(i-1),\, w(i-1)+1,\,\dots,\, wi-1\}
\qquad (i\ge 1).
\]
Let $\phi=\phi_{\mathrm{med}}$ denote the unique median representative of each grain.
Then the coarse cell addition satisfies
\[
G_{\pi,i}\boxplus_{\pi,\phi} G_{\pi,j}=G_{\pi,i+j-1}
\qquad (i,j\ge 1),
\]
and hence is associative:
\[
(G_{\pi,i}\boxplus_{\pi,\phi} G_{\pi,j})\boxplus_{\pi,\phi} G_{\pi,k}
=
G_{\pi,i}\boxplus_{\pi,\phi}(G_{\pi,j}\boxplus_{\pi,\phi} G_{\pi,k}).
\]
\end{proposition}

\begin{proof}
Since each grain has odd width $w=2m+1$, its unique median is
\[
\phi(G_{\pi,i})=w(i-1)+m.
\]
Therefore
\[
\phi(G_{\pi,i})+\phi(G_{\pi,j})
=
w(i-1)+m+w(j-1)+m
=
w(i+j-2)+2m.
\]
Because $2m=w-1$, this becomes
\[
\phi(G_{\pi,i})+\phi(G_{\pi,j})
=
w(i+j-2)+(w-1)
=
w(i+j-1)-1.
\]
But $w(i+j-1)-1$ is exactly the maximum element of the grain $G_{\pi,i+j-1}$, since
\[
G_{\pi,i+j-1}
=
\{w(i+j-2),\,\dots,\,w(i+j-1)-1\}.
\]
Hence
\[
G_{\pi,i}\boxplus_{\pi,\phi} G_{\pi,j}
=
\psi_\pi(\phi(G_{\pi,i})+\phi(G_{\pi,j}))
=
G_{\pi,i+j-1}.
\]

Therefore
\[
(G_{\pi,i}\boxplus_{\pi,\phi} G_{\pi,j})\boxplus_{\pi,\phi} G_{\pi,k}
=
G_{\pi,i+j-1}\boxplus_{\pi,\phi} G_{\pi,k}
=
G_{\pi,i+j+k-2},
\]
and similarly
\[
G_{\pi,i}\boxplus_{\pi,\phi}(G_{\pi,j}\boxplus_{\pi,\phi} G_{\pi,k})
=
G_{\pi,i}\boxplus_{\pi,\phi} G_{\pi,j+k-1}
=
G_{\pi,i+j+k-2}.
\]
Thus the two bracketings agree.
\end{proof}

\begin{remark}
The proposition above should be understood as an exceptional case rather than the
generic behavior of coarse addition. In most partitions, especially those with
nonuniform grain widths or asymmetric representatives, the repeated projection to
grain representatives destroys associativity.
\end{remark}

\section{Application to the St.\ Petersburg Paradox}\label{sec6}

\subsection{The Classical Series and the Present Strategy}

The classical St.\ Petersburg game \cite{Bernoulli1738} is defined by the payoff rule
\[
\Pr(X=2^{n-1})=2^{-n}
\qquad (n=1,2,3,\dots).
\]
Under ordinary expectation, one obtains
\[
\mathbb E[X]
=
\sum_{n=1}^{\infty} 2^{-n}\cdot 2^{n-1}
=
\sum_{n=1}^{\infty} \frac12,
\]
which diverges to $+\infty$.

The present framework does not attempt to replace probability theory itself.
Rather, it asks how such an infinite sequence of equal expected increments may be
aggregated under a coarse-grained addition rule. Since our basic setup is formulated
on $U=\mathbb N_0$, we work with the rescaled sequence
\[
b_n:=2\cdot \Bigl(2^{-n}\cdot 2^{n-1}\Bigr)=1
\qquad (n\in\mathbb N).
\]
Thus, instead of the ordinary divergent series
\[
\frac12+\frac12+\frac12+\cdots,
\]
we consider the equivalent integer-valued constant sequence
\[
1,1,1,\dots
\]
and study its left-associative coarse partial sums.

This rescaling is adopted only to place the constant expected increment sequence on the discrete domain $\mathbb{N}_0$, not to redefine the classical expectation itself. The question is not
whether the classical expectation converges in the ordinary sense---it does not---but
whether a coarse-grained aggregation process may become inert after finitely many steps.

\subsection{An Explicit Partition Yielding Inertness}

We now exhibit a countable discrete coarse-grained partition under which the constant
sequence $(1,1,1,\dots)$ becomes inert. The triangular partition is introduced here only as an explicit and mathematically convenient example. We do not claim that it is the uniquely correct partition for modeling actual human cognition.

\begin{definition}[Triangular partition]
Define a countable discrete coarse-grained partition
\[
\pi^\triangle = \{G_n^\triangle\}_{n\ge 1}
\]
of $\mathbb N_0$ by
\[
G_n^\triangle
:=
\{T_{n-1},\,T_{n-1}+1,\,\dots,\,T_n-1\},
\]
where
\[
T_n:=\frac{n(n+1)}{2}
\qquad (n\ge 0)
\]
is the $n$th triangular number, with $T_0:=0$.
Equivalently,
\[
|G_n^\triangle|=n
\qquad (n\ge 1).
\]
\end{definition}

Thus the first few grains are
\[
G_1^\triangle=\{0\},\qquad
G_2^\triangle=\{1,2\},\qquad
G_3^\triangle=\{3,4,5\},\qquad
G_4^\triangle=\{6,7,8,9\},
\]
\[
G_5^\triangle=\{10,11,12,13,14\},\qquad \dots
\]

We now show that, for this partition, inertness arises under a suitable choice of internal representative map. Let us choose the minimum representative
\[
\phi_{\min}(G_n^\triangle)=\min G_n^\triangle=T_{n-1}.
\]
Then the absorption margin is
\[
\mu_{\phi_{\min}}(G_n^\triangle)
=
(T_n-1)-T_{n-1}
=
n-1.
\]

Since the repeated increment is always $1$, it follows that every grain
$G_n^\triangle$ with $n\ge 2$ satisfies
\[
1\le \mu_{\phi_{\min}}(G_n^\triangle).
\]
Hence every such grain absorbs the singleton increment represented by~$1$.

\begin{proposition}[Inertness of the rescaled St.\ Petersburg increment sequence]
Let
\[
x_n=1
\qquad (n\in\mathbb N),
\]
and let $\pi^\triangle$ be the triangular partition above with representative map
$\phi_{\min}$. Then the left-associative coarse partial sums
\[
S_1^{\mathrm{rep}}:=1,
\qquad
S_n^{\mathrm{rep}}:=S_{n-1}^{\mathrm{rep}}\oplus_{\pi^\triangle,\phi_{\min}} 1
\quad (n\ge 2)
\]
are inert.
More precisely,
\[
S_n^{\mathrm{rep}}=1
\qquad\text{for all }n\ge 1.
\]
\end{proposition}

\begin{proof}
Since $1\in G_2^\triangle=\{1,2\}$, we have
\[
\psi_{\pi^\triangle}(1)=G_2^\triangle.
\]
Also,
\[
\phi_{\min}(G_2^\triangle)=1,
\qquad
\mu_{\phi_{\min}}(G_2^\triangle)=2-1=1.
\]
Therefore the constant increment grain $\psi_{\pi^\triangle}(1)=G_2^\triangle$
satisfies
\[
\phi_{\min}(G_2^\triangle)=1\le \mu_{\phi_{\min}}(G_2^\triangle)=1.
\]
By the absorption criterion via margin,
\[
G_2^\triangle\boxplus_{\pi^\triangle,\phi_{\min}} G_2^\triangle
=
G_2^\triangle.
\]
Hence
\[
S_1^{\mathrm{cell}}=G_2^\triangle
\]
and every later increment is absorbed by the same grain. By the margin-based
sufficient condition for inertness, the process is inert at~$G_2^\triangle$.
Consequently,
\[
S_n^{\mathrm{rep}}=\phi_{\min}(G_2^\triangle)=1
\qquad\text{for all }n\ge 1.
\]
\end{proof}

\subsection{Interpretation}

The significance of the preceding proposition is not that the classical expectation
has somehow been made finite in the ordinary probabilistic sense. Rather, it shows
that once aggregation is performed through a coarse-grained arithmetic, a sequence of
identical positive increments may become dynamically ineffective after finitely many
steps, or even immediately.

In the present example, the expected increment at each stage is constant. Under
ordinary addition, this yields divergence. Under coarse addition, however, the running
total is repeatedly projected back to the representative of its current grain. If the
grain has sufficient absorption margin relative to the incoming increment, then the
increment ceases to change the coarse state. The process therefore becomes inert.

This suggests the following interpretation. The St.\ Petersburg paradox relies on the
fact that classical expectation treats every additional expected contribution, however
small in practical cognitive terms, as cumulatively effective. By contrast, coarse
addition models a bounded mode of aggregation in which sufficiently small repeated
increments may eventually fail to alter the perceived total.

It is important to distinguish the present interpretation from the more familiar use
of discounting. In game theory, discounting is often introduced in order to make an
infinite payoff stream convergent by assigning progressively smaller weights to later
payoffs. In that respect, discounting and coarse addition may appear similar, since
both modify the treatment of an otherwise divergent infinite sequence. However, the
two approaches are grounded in fundamentally different modeling assumptions.

Discounting is intended to capture a temporal feature of evaluation: payoffs received
in the more distant future are assigned less significance than those received sooner.
Coarse addition, by contrast, does not model temporal remoteness. Instead, it models
limited sensitivity to increments that are small relative to an already accumulated
total. On this view, the issue is not that a payoff occurs later, but that, once the
current coarse state is sufficiently large, an additional increment may fail to alter
it.

This difference may be illustrated informally as follows. In a game offering an
additional payoff of \$10{,}000 upon success, that increment may be highly significant
for a player with no wealth, but practically negligible for a player who already
possesses billions of dollars. Coarse addition is designed to formalize this kind of
bounded responsiveness to further gains relative to an existing total.

One advantage of this perspective is that it allows a certain analysis of the
St.\ Petersburg paradox even when temporal considerations are set aside. The present
framework therefore isolates a possible source of bounded evaluation that is distinct
from the temporal assumptions built into discounting.

The present result should be interpreted with caution.

\begin{enumerate}
\item It does \emph{not} show that the St.\ Petersburg paradox is ``solved'' in the
sense of standard decision theory. Ordinary expectation remains divergent.
\item The result depends on the chosen partition and representative map.
Different coarse-grained structures may yield different outcomes, including non-inert
behavior.
\item What has been shown is narrower but still conceptually significant:
there exist coarse-grained aggregation schemes under which the sequence of expected
increments associated with the St.\ Petersburg game becomes inert. This supports the
idea that bounded or coarse numerical cognition may respond to divergent reward
structures differently from classical exact arithmetic.
\end{enumerate}

Accordingly, the relevance of the present framework is heuristic and structural.
It isolates one mathematically explicit mechanism---absorption under coarse
addition---through which an infinite reward structure may fail to generate an
ever-increasing perceived total.

\section{Conclusion}\label{sec7}

This paper has introduced a coarse-grained framework for studying how divergent reward structures may be aggregated under bounded numerical resolution. Rather than modifying utility functions, introducing discount factors, or appealing to extended number systems, the present approach alters the aggregation rule itself. The key idea is to partition the underlying scale into ordered grains, assign to each grain an internal representative, and define addition through repeated projection to those representatives.

Within this framework, we identified several structural features of coarse addition. First, coarse addition gives rise to absorption: under suitable conditions, the addition of one grain to another leaves the latter unchanged. Second, repeated absorption yields inertness, namely eventual stabilization of a left-associative coarse addition process at a fixed grain and hence at a fixed representative value. Third, coarse addition is generally non-associative, although special associative cases may occur under highly regular partition structures. Taken together, these results show that coarse addition is not merely an approximation to ordinary addition, but a distinct algebraic mechanism with its own characteristic behavior.

The application to the St.\ Petersburg paradox should be interpreted in this restricted but meaningful sense. The paper does not show that the paradox is ``solved'' within standard decision theory, nor does it render the classical expected value finite in the ordinary probabilistic sense. What it does show is that there exist coarse-grained aggregation schemes under which the sequence of equal expected increments associated with the St.\ Petersburg game becomes inert after coarse re-aggregation. This supports the heuristic claim that an agent with limited cognitive precision may process a divergent reward structure differently from an agent employing exact arithmetic.

Accordingly, the contribution of the paper is primarily structural and heuristic. It isolates one explicit mathematical mechanism---absorption under coarse addition---through which an infinite sequence of positive increments may cease to alter the perceived total. In this respect, the framework may be relevant not only to decision-theoretic paradoxes, but also to broader questions concerning bounded rationality, coarse numerical cognition, and coarse-grained models of aggregation.

Several limitations remain. The present study is confined to countable discrete partitions of \(N_0\) with internal representatives, and all infinite sums are treated in a left-associative manner. Moreover, the St.\ Petersburg application depends on a particular choice of partition and representative map; different coarse structures may produce different outcomes, including non-inert behavior. For these reasons, the present framework should be understood not as a universal account of human valuation, but as a mathematically explicit model of one possible mode of coarse-grained aggregation.

Future work may proceed in several directions. One is to classify more systematically which partitions and representative maps generate inertness or associativity. Another is to investigate how coarse addition relates to other forms of bounded aggregation in decision theory and behavioral modeling. A further direction is to examine whether similar coarse-grained operators can be used to study aggregation processes in settings where exact numerical information is compressed into cognitively manageable categories.

In sum, the main lesson of the paper is modest but, we believe, conceptually important: once aggregation itself is made coarse, divergence in the classical sense need not translate into unbounded growth at the coarse level. Coarse addition thus opens a new way of thinking about how infinite reward structures may appear under bounded modes of representation and computation.

\section*{Declarations}

\textbf{Funding} \\
This research received no specific grant from any funding agency in the public, commercial, or not-for-profit sectors.

\noindent
\textbf{Conflicts of Interest} \\
The author declares no conflict of interest.

\noindent
\textbf{Ethical Approval} \\
This article does not contain any studies with human participants or animals performed by any of the authors.

\noindent
\textbf{Data Availability} \\
No empirical data was used in this study.

\bibliography{references}

\end{document}